\documentclass[english,aps, twocolumn]{revtex4}
\usepackage[T1]{fontenc}
\usepackage[latin9]{inputenc}
\usepackage[letterpaper]{geometry}
\usepackage{color}
\usepackage{amstext}
\usepackage{graphicx}
\usepackage{esint}

\makeatletter
\@ifundefined{textcolor}{}
{%
 \definecolor{BLACK}{gray}{0}
 \definecolor{WHITE}{gray}{1}
 \definecolor{RED}{rgb}{1,0,0}
 \definecolor{GREEN}{rgb}{0,1,0}
 \definecolor{BLUE}{rgb}{0,0,1}
 \definecolor{CYAN}{cmyk}{1,0,0,0}
 \definecolor{MAGENTA}{cmyk}{0,1,0,0}
 \definecolor{YELLOW}{cmyk}{0,0,1,0}
 }

\makeatother

\usepackage{babel}

\begin{document}

\title{Aperiodic conductivity oscillations in quasi-ballistic graphene heterojunctions}

\author{Milan Begliarbekov$^{1}$, Onejae Sul$^{2}$, Nan Ai$^{1}$, Eui-Hyeok
Yang$^{2}$, Stefan Strauf$^{1}$}

\email{strauf@stevens.edu}

\affiliation{$^{1}$Department of Physics and Engineering Physics, Stevens Institute
of Technology, Hoboken NJ 07030, USA}

\affiliation{$^{2}$Department of Mechanical Engineering, Stevens Institute of
Technology, Hoboken NJ 07030, USA}
\begin{abstract}
We observe conductivity oscillations with aperiodic spacing to only
one side of the tunneling current in a dual-gated graphene field effect
transistor with an n-p-n type potential barrier. The spacing and width
of these oscillatoins were found to be inconsistent with pure Farbry-Perot-type
interferences, but are in quantitative agreement with theoretical
predictions that attribute them to resonant tunneling through quasi-bound
impurity states. This observation may be understood as another signature
of Klein tunneling in graphene heterojunctions and is of importance
for future development and modeling of graphene based nanoelectronic
devices. 
\end{abstract}
\maketitle
Graphene is a two-dimensional monolayer of carbon atoms that results
in a zero-gap semiconductor with outstanding electronic \cite{Geim}
and thermal properties \cite{Balandin}. Unlike in most conventional
semiconductors, charge carriers in graphene obey the Dirac equation
and are capable of ballistic \cite{Beenakker08,Stander09,Gorbachev08,Williams07}
and coherent transport \cite{Russo08,Young09}, and Veselago lensing
\cite{Cheianov07}. One of the most promising devices for applications
in nanoelectronics is the graphene field effect transistor (GFET),
which was shown to be capable of ultra high frequency (100 GHz) operation
\cite{Lin10}. 

Locally gated GFETs give rise to more complex architectures such as
n-p-n heterojunctions, which can be realized without physically doping
the underlying material. In the conventional transport regime the
tunneling current across a potential barrier decreases with increasing
barrier energy. In contrast, the chiral Fermions in graphene have
been predicted to tunnel through potential barriers with near unitary
probability \cite{Beenakker08,Katsnelson}. As a result, an increasing
tunneling current with increasing barrier energy is expected, analogous
to Klein tunneling in quantum electrodynamics. Furthermore, top gated
graphene heterojunctions form a Fabry-Perot (FP) type cavity for electron
waves, which undergo oscillations in the top gated region due to multiple
reflections within the barrier. A $\pi$-phase shift of the FP oscillations
in a magnetic field was recently observed and understood to be the
signature of Klein tunneling in graphene heterojunctions \cite{Young09,Gorbachev08,Shutov08}.
Recently, Rossi et al. predicted that the residual impurity concentration
in partially disordered junctions gives rise to a non-negligible scattering
potential $V_{sc}$, which causes broad and aperiodic conductivity
oscillatio\textcolor{black}{ns, which might be understood as another
signature of Klein tunneling \cite{Rossi10}.}

Here we report the experimental observation of aperiodic conductivity
oscillations in a quasi-ballistic GFET. We analyze the spacing and
the width of these oscillations, and find that both are better explained
by the resonant tunneling model, and cannot be attributed solely to
Fabry-Perot oscillations. We further analyze the disorder potential
introduced by charged impurities and phonons at elevated temperatures. 

A schematic of the dual-gated GFET and the electrical biasing scheme
are shown in Fig 1a. 

\begin{figure}
\includegraphics[scale=0.38]{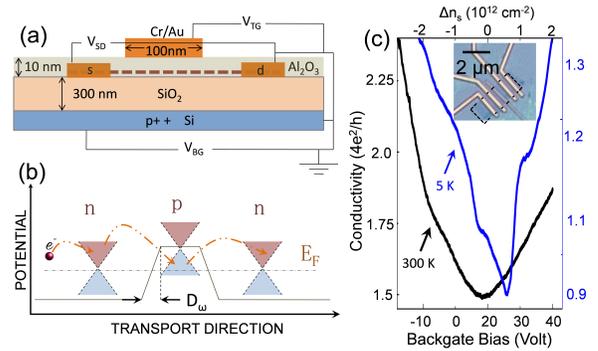}

\caption{(a) Schematic of device geometry and contacting scheme. (b) Potential
profile created by the global backgate and the local top gate as a
function of the device length (transport direction) creating an n-p-n
junction. (c) Room temperature (black line) and cryogenic (blue line)
conductivities of the device. Inset: Optical micrograph of the GFET
- the graphene flake has been outlined for clarity.}

\end{figure}

The graphene flakes were prepared by micromechanical exfoliation of
natural graphite onto a $\mathrm{\textrm{p}^{+\!+}}$ Si wafer with
a thermally grown $\mathrm{\textrm{300\textrm{ }\textrm{nm}\textrm{ }Si\ensuremath{\textrm{O}_{2}}}}$
dielectric. The flakes were identified as being single layer from
their characteristic Raman spectrum \cite{Begliarbekov10,Ferrari06}.
Electron beam lithography was utilized to pattern the graphene flake
and to define electrical Cr/Au contacts. A 10 nm $\mathrm{Al_{2}O_{3}}$
layer was then evaporated to provide a gate oxide for the subsequent
deposition of a 100 nm wide top-gate. An optical micrograph of a device
similar to the one used in these experiments is shown in the inset
to Fig. 1c. The top gate was used to apply a local electrostatic potential,
thereby creating an n-p-n junction as shown in Fig. 1b, with a partially
graded junction of width $D_{\omega}$. 

The visibility of conductivity oscillations depends strongly on the
underlying mobility, coherence length, and the disorder potential.
We first characterize the relevant transport parameters in our GFET
device by grounding the top gate to the drain electrode. Consequently,
carrier transport in the resultant structure is similar to transport
in a graphene nanoribbon (GNR). A constant 100 mV source drain bias
$V_{sd}$ was applied while the global backgate bias $V_{bg}$ was
varied from $\mathrm{\textrm{-40 V}}$ to $\mathrm{\textrm{+40 V}}.$
The relatively high backgate bias was required since a 300 nm thick
dielectric is necessary to provide good optical contrast for the purpose
of locating graphene flakes. The mobility can be estimated from the
data in Fig.1c using $\mu=\left(en\rho\right)^{-1}$, where $n=C_{ox}(V_{bg}-V_{Dirac})/e$,
$V{}_{Dirac}$ the voltage at the charge neutrality point, and $C_{ox}=115\:\mathrm{aF/\mu m^{2}}$the
oxide capacitance \cite{Stander09,Tan07,Huard07}. At carrier densities
of $1-2\times10^{12}\textrm{ cm}^{-2}$, corresponding to back gate
voltages of 30-40 V, we estimate a room temperature (RT) mobility
of $1120\textrm{ }\mathrm{cm^{2}V^{-1}s^{-1}}$ and a cryogenic (5
K) mobility of $3300\mathrm{\textrm{ }cm^{2}V^{-1}s^{-1}.}$ The corresponding
ballistic mean free paths of $l_{e}\cong\mathrm{50\textrm{ }nm}$
at RT and $l_{e}\cong\mathrm{110\textrm{ }nm}$ at 5K were estimated
from the scattering time \cite{Tan07}. Note that the $\sigma-V_{bg}$
curve in Fig. 1c shows some kinks that most likely originate from
tunneling through trapped states which originate from Fermi-level
pinning of the local potential at the impurities. These kinks occur
on both sides of the global conductivity minimum, and their position
changes with each cooldown event. 

Furthermore, the 2D sheet carrier density $\Delta n_{s}$ was estimated
from the geometry of the device and known material properties using
$\Delta n_{s}=\epsilon_{o}\epsilon_{r}(V_{bg}-V_{Dirac})/ed_{ox}$,
where $d{}_{ox}=300\:\:\mathrm{nm}$ is the oxide thickness, $\varepsilon_{r}$
is the dielectric constant of $\textrm{Si\ensuremath{\textrm{O}_{2}}}$.
Similarly, the corresponding top-gate carrier density $\Delta n_{TG}$
can be calculated from the sheet carrier density $\Delta n_{s}$ \cite{Stander09}. 

Unlike top and back gate potentials, $\Delta n_{s}$ provides unambiguous
information about the underlying transport since $\Delta n_{s}$ is
zero at the charge neutrality point, which is not necessarily located
at $V_{bg}=0$. The location of the charge neutrality point away from
$V_{bg}=0$ originates from the presence of charged impurities, which
contribute to the conductivity $\sigma_{ci}$ according to \cite{Chen08,Nomura07}:
$\sigma_{ci}\left(n\right)=C_{imp}\left|\Delta n_{s}/n_{imp}\right|,$
where $C_{imp}=5\times10^{15}\textrm{ \ensuremath{V^{-1}s^{-1}}}$
is a constant related to the screened Coulomb potential \cite{Hwang07},
and $n_{imp}$ is the impurity density. From this equation we find
$n_{imp}=6\times10^{11}$ $\textrm{c\ensuremath{m^{-2}}}$ in our
device, the knowledge of which is crucial in identifying the transport
regime, and correlating it to the visibility of conductivity oscillations.
We further calculate the $\beta$ parameter, given by $\beta=n'n_{i}^{-3/2}$,
where $n'$ is the slope of the density profile around zero density,
and $n_{i}\equiv e/\mu h$, which differentiates between diffusive
and ballistic regimes \cite{Folger08}. Values of $\beta\ll1$ indicate
purely diffusive transport, whereas values of $\beta\gg1$ are characteristic
of the ballistic regime. In our device $\beta=3.7$ at 5 K, which
is indicative of the quasi-ballistic regime were both ballistic and
diffusive transport contributes to the conductivity. 

Following the initial characterization of the GNR, the main results
have been achieved by applying a bias $V_{TG}$ to the top gate thereby
creating the electrostatic potential shown in Fig. 1b. Using this
configuration, we observed an increasing tunneling current with increasing
barrier height in the vicinity of the charge neutrality point, as
shown in Fig. 2a. %
\begin{figure}
\includegraphics[scale=0.41]{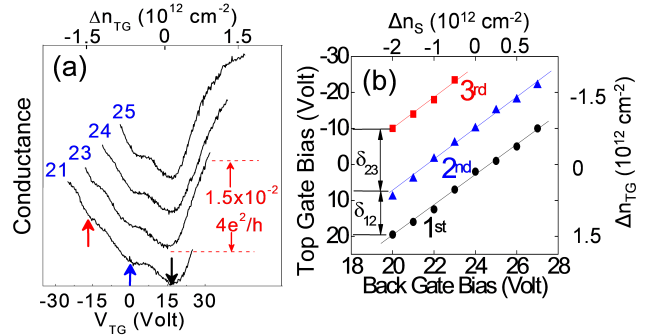}

\caption{(a) Conductivity as a function of back and top gate bias showing conductivity
oscillations characteristic of Klein tunneling. The lines mark conductivity
minima. (b) Positions of the peak minima as a function of top and
back gate bias. Data recorded at 5 K.}

\end{figure}
 In addition, the onset of up to three conductivity oscillation minima
(resistivity maxima) \textcolor{black}{are visible in an aperiodic
spacing to only one side of the global conductivity minimum $\Delta n_{tg}=0$. }

\textcolor{black}{Similar conductance oscillations have been previously
observed \cite{Velasco09,Huard07,Gorbachev08}, and analyzed in the
context of FP oscillations. The oscillations observed in our device
cannot be explained solely by the FP model. In the FP model, where
the k-vector is affected by the geometric boundary, the magnitude
of the spacing between successive oscillations $\delta_{ij}$ can
be approximated by the condition $k_{F}\left(2L\right)=2\pi$. Consequently
the peak spacing becomes $\delta_{ij}=4\sqrt{\pi n_{2}}/L_{c}$, where
$L_{c}\cong L_{tg}+2d$, and is thus constant \cite{Velasco09}, which
would be on the order of $5\times10^{10}\textrm{ cm}^{-2}$ in our
device. However, in the presence of a scattering potential $V_{sc}$
due to impurity states, the phase shift of the interference $\theta_{WKB}$
is given by $\theta_{WKB}=-\int V_{sc}\left(x',y\right)dx'$, where
the scattering potential $V_{sc}$ is proportional to $V_{sc}\sim\left[V_{d}\left(\mathbf{r}\right)+V_{tg}\left(\mathbf{r}\right)+\frac{1}{2}\int d^{2}r'\frac{n\left(\mathbf{r}'\right)}{\left|\mathbf{r}-\mathbf{r}'\right|}\right]$
\cite{Rossi10}. Using the above expression for the impurity potential
Rossi et al. obtain peak spacings of $\delta_{12}=0.85\textrm{\ensuremath{\times}1\ensuremath{0^{12}} cm}^{-2}$
and $\delta_{23}=1.0\textrm{ \ensuremath{\times}1\ensuremath{0^{12}} cm}^{-2}$
at an impurity concentration $n_{imp}=5\times10^{11}\textrm{ cm}^{-2}$
\cite{Rossi10}. In our device $\delta_{12}=0.8\times10^{12}\textrm{ cm}^{-2}$
and $\delta_{23}=1.1\times10^{12}\textrm{ cm}^{-2}$ at the estimated
impurity concentration of $n_{imp}=6\times10^{11}\textrm{ cm}^{-2}$.}

\textcolor{black}{Furthermore, the observed aperiodic spacing, i.e.,
$\delta_{12}\neq\delta_{23}$ as shown in Fig. 2b, can be accounted
for by the fact that $V_{sc}$ is a function of topgate bias as well
as the carrier concentration inside the junction. This feature cannot
be explained solely by the FP model but it is present in the self-consistent
simulations using the above expression for the scattering potential.
In addition, the width of our oscillations is larger than the theoretically
predicted width of FP oscillations \cite{Shutov08}, and the experimentally
measured width in cleaner devices \cite{Young09,Gorbachev08}. In
contrast, in the presence of $V_{sc}$, the visibility of these oscillations
is strongest at low impurity concentrations and decreases with increasing
impurity density, but are predicted to be still visible at our experimental
values of $n_{imp}=6\times10^{11}$ $\mathrm{cm^{-2}}$, while both
the theoretical and experimental width are estimated to be about $0.4\times10^{12}\textrm{ cm}^{-2}$,
and thus comparable. Consequently, the magnitude of oscillation spacing
and width in our device is best explained by resonant tunneling through
quasi-bound impurities. }

Finally, we analyze the temperature dependence of the conductivity
oscillations to study the influence of the degradation of the ballistic
mean free path and polar optical phonons on the observed oscillations.
The magnitude of oscillations diminishes at higher temperatures as
shown in Fig.3a. %
\begin{figure}
\includegraphics[scale=0.4]{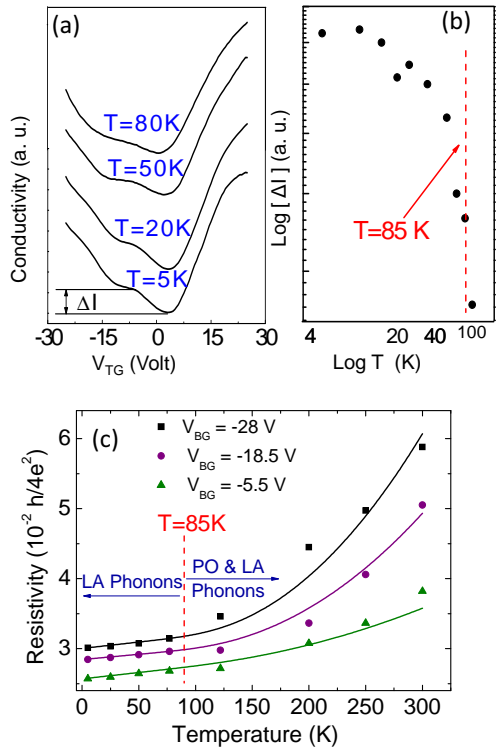}

\caption{(a) Temperature dependence of the conductivity oscillations. (b) Amplitude
$\Delta\mathrm{I}$ of the first oscillation with respect to the minimum
between the first and second peak. (c) Temperature dependence of the
resistivity as a function of several back-gate biases.}

\end{figure}
 Figure 3b shows a log-log plot of the amplitude of the first resistance
oscillation peak $\Delta$I as a function of temperature, which vanishes
at about 85 K. To understand what happens at this particular temperature
we recorded the temperature dependence of the GFET resistivity with
the top-gate grounded as shown in Fig. 3c. The observed exponential
increase in resistivity is consistent with previous investigations
\cite{Chen08,Meric}. The fitted solid curves in Fig. 3c correspond
to the phonon contribution to the resistivity $\rho_{ph}$ as given
by $\rho_{ph}=\rho_{0}\left(V_{BG}\right)+\rho_{LA}\left(T\right)\mathbf{+}\rho_{PO}\left(V_{BG},T\right)$,
where $\rho_{0}\left(V_{BG}\right)$ is the residual resistivity,
and $\rho_{LA}$ and $\rho_{PO}$ are due to acoustic and polar optical
phonons, respectively \cite{Chen08}. The only free parameter in $\rho_{LA}$
is the acoustic deformation potential $D_{A}$, which we extract from
the linear part in Fig. 3c to be $D_{A}=15\pm3$ eV. The contribution
to the total phonon resistivity due to $\rho_{PO}$ was analyzed following
Ref. \cite{Fratini08}, and is in good agreement with our data (see
fit in Fig 3c). 

Thus, the underlying scattering mechanism most likely originates from
polar phonon injection from the underlying $\textrm{Si\ensuremath{\textrm{O}_{2}}}$,
which set in above 85 K. These phonons cause carrier scattering and
thereby degrade the ballistic mean free path. As a consequence, transport
through the barrier becomes purely diffusive, since $l_{e}\cong\mathrm{110\textrm{ }nm}$
at 5 K degraded to values smaller than the top gate length of 100
nm which defines the n-p-n junction. Therefore, the conductance oscillations
vanish at about 85 K when the GFET transitions from quasi ballistic
to diffusive transport through the barrier due to the onset of polar
optical phonons.

In summary, we fabricated an GFET with an n-p-n type potential barrier
and observed aperiodic conductivity oscillations in the quasi-ballistic
regime ($\beta=3.7$). The peak spacing cannot be explained solely
by the FP model, but is correctly predicted when resonant tunneling
through impurity states is taken into account, in agreement with recent
theoretical predictions. The observation of resonant tunneling through
impurity states may be understood as another signature of Klein tunneling
in graphene heterojunctions and is of importance for future development
of high performance GFETs. 
\begin{acknowledgments}
We thank Kitu Kumar, Steve Tsai, and Anderson Tsai for assistance
with sample preparation. Partial financial support was provided by
the NSF GK-12 Grant No. DGE-0742462 and by AFOSR, award No. FA9550-08-1-013.
We thank the Center for Functional Nanomaterials of the Brookhaven
National Laboratory, contract DE-AC02-98CH10886, for the use of their
clean rooms.\end{acknowledgments}


\begin{thebibliography}{24}
\bibitem{Geim}A. K. Geim and K. S. Novoselov, Nature Materials \textbf{7},
183 (2007).

\bibitem{Balandin}A. A. Balandin, S. Ghosh, W. Bao, I. Calizo, D.
Teweldebrhan, F. Miao, and C. N. Lau, Nano Letters \textbf{8}, 902
(2008).

\bibitem{Stander09}N. Stander, B. Huard, and D. Goldhaber-Gordon,
Phys. Rev. Lett. \textbf{102,} 026807 (2009).

\bibitem{Williams07}J. R. Williams, L. DiCarlo, and C. M. Marcus,
Science \textbf{317}, 638 (2007).

\bibitem{Gorbachev08}R. V. Gorbachev, A. S. Mayorov, A. K. Savchenko,
D. W. Horsell, and F. Guinea, Nano Lett. \textbf{8}, 1995 (2008).

\bibitem{Beenakker08}C. W. J. Beenakker, Reviews of Modern Physics
\textbf{80}, 1337 (2008).

\bibitem{Young09}A. F. Young and P. Kim, Nature Physics \textbf{5},
222 (2009).

\bibitem{Russo08}S. Russo, J. B. Oostinga, D. Wehenkel, H.t B. Heersche,
S. S. Sobhani, L. M. K. Vandersypen, and A. F. Morpurgo, Phys. Rev.
B \textbf{77}, 085413 (2008).

\bibitem{Cheianov07}V. V. Cheianov, V. Fal'ko, and B. L. Altshuler,
Science \textbf{315}, 1252 (2007).

\bibitem{Lin10}Y.-M. Lin, C. Dimitrakopoulos, K. A. Jenkins, D. B.
Farmer, H.-Y. Chiu, A. Grill, Ph. Avouris., Science \textbf{327},
662 (2010).

\bibitem{Katsnelson}M. I. Katsnelson, K. S. Novoselov, and A. K.
Geim, Nat. Phys. \textbf{2}, 620 (2006).

\bibitem[12]{Shutov08}V. Shytov, M. S. Rudner, and L. S. Levitov,
Phys. Rev. Lett. \textbf{101}, 186806 (2008).

\bibitem[13]{Rossi10}E. Rossi, J. H. Bardarson, P. W. Brouwer, and
S. D. Sarma, Phys. Rev. B \textbf{81}, 121408(R) (2010).

\bibitem[14]{Begliarbekov10}M. Begliarbekov, O. Sul, S. Kalliakos,
E.-H. Yang, and S. Strauf, Appl. Phys. Lett. \textbf{97}, 031908 (2010).

\bibitem[15]{Ferrari06}A. C. Ferrari, J. C. Meyer, V. Scardaci, C.
Casiraghi, M. Lazzeri, F. Mauri, S. Piscanec, D. Jiang, K. S. Novoselov,
S. Roth, and A. K. Geim, Phys. Rev. Lett., \textbf{97}, 187401 (2006).

\bibitem[16]{Tan07}Y.-W. Tan1, Y. Zhang1, K. Bolotin, Y. Zhao, S.
Adam, E. H. Hwang, S. Das Sarma, H. L. Stormer, and P. Kim, Phys.
Rev. Lett. \textbf{99}, 246803 (2007).

\bibitem[17]{Huard07}B. Huard, J. A. Sulpizio, N. Stander, K. Todd,
B. Yang, and D. Goldhaber-Gordon, Phys. Rev. Lett., \textbf{98}, 236803
(2007).

\bibitem[18]{Chen08}J.-H. Chen, C. Jang, M. Ishigami, S. Xiao, E.
D. Williams, and M. S. Fuhrer, Sol. State Comm. \textbf{149}, 1080
(2008).

\bibitem[19]{Nomura07}K. Nomura and A. H. MacDonald, Phys. Rev. Lett.
\textbf{98}, 076602 (2007).

\bibitem[20]{Hwang07}E. H. Hwang, S. Adam, and S. D. Sarma, Phys.
Rev. Lett. \textbf{98}, 186806 (2007).

\bibitem[21]{Folger08}M. M. Fogler, D. S. Novikov, L. I. Glazman,
and B. I. Shklovshii, Phys. Rev. B \textbf{77}, 075420 (2008).

\bibitem[22]{Velasco09}J. Velasko, G. Liu, W. Bao, and C. N. Lau,
New J. Phys. \textbf{11}, 095008 (2009).

\bibitem[23]{Meric}I. Meric, M. Y. Han, A. F. Young, B. Ozyilmaz,
P. Kim, and K. L. Shepard, Nat. Nano. \textbf{3}, 654 (2008).

\bibitem[24]{Fratini08}S. Fratini and F. Guinea, Phys. Rev. B \textbf{77},
195415 (2008).
\end{thebibliography}
\end{document}